\documentclass[runningheads]{llncs}
\usepackage[T1]{fontenc}
\usepackage{xcolor}
\usepackage{threeparttable}
\usepackage{multirow}
\usepackage{subcaption}
\usepackage{graphics}
\usepackage{tikz}
\usetikzlibrary{arrows,automata,positioning}
\usepackage{tikz}
\usepackage{bibnames}
\usepackage[linesnumbered,ruled,vlined]{algorithm2e}
\usepackage{diagbox}
\begin{document}
\title{Cyber Physical Aquaponic System (CyPhA): a CPS Testbed}
%
%
\author{Anand Agrawal
\and
Praneeta Maganti
\and
Rajib Ranjan Maiti
}
\authorrunning{A Agrawal et al.}
%
%

\institute{Birla Institute of Technology and Science Pilani, Hyderbad Campus, 500078, IND }
\maketitle              
\begin{abstract}
Aquaponics system promises a sustainable urban development and food production by combining vegetable and fish farming in a single water loop. 
However, traditional aquaponics suffers from a significant amount of manual intervention with regard to decision-making in the water circulation and water quality control. 
In this work, we design, build and deploy a laboratory-scale real aquaponics system by considering this system as a cyber physical system, and we call it as \emph{C}yber \emph{Ph}ysical \emph{A}quaponics (CyPhA) system.  
The design of our CyPhA system has five stages, Stage-1 contains a vertical vegetable farming unit, Stage-2 contains fish farming unit, Stage-3 contains natural nitrification system, Stage-4 contains bio-filtration system and Stage-5 contains water accumulation and release system. 
Water transfer from one stage to the next is done using water pumps, and oxygen mixing in the water in any stage is achieved using aeration pumps. 
CyPhA system uses sensors for pH, dissolved oxygen (DO), total dissolved solid (TDS), water temperature, air temperature and humidity.
A critical level of any of the water parameters in any stage is indicated using a LED-based alert indicator. 
Sensor data and actuator control commands among the stage-wise edge devices and the CyPhA Controller are exchanged over Message Queue Telemetry Transport (MQTT) protocol.
Overall, CyPhA system is housed within an area of about 80 sq. ft. 
We have been successfully operating CyPhA system for the last 75 days and maintaining a good quality of water for both fish \footnote{We have used fish water, instead of fish, collected from a local fish vendor due to ethical issues of utilizing animal at present.} and vegetable farming units.

\end{abstract}
\section{Introduction}
\label{sec:intro}
Aquaponics market is growing at a commendable rate of about [12.6, 14.1]\% globally in recent times, and it is expected to grow in a similar rate till 2030 according to certain market studies \cite{openpr}, \cite{marketresfuture}. 
With the increasing population in metropolitan cities, aquaponics system that combines vegetable and fish farming together is a sustainable food production ecosystem that produces leafy vegetables and edible fish within a much smaller land area, compared to isolated traditional ground based vegetable or fish farming. 
More importantly, this system achieves an excellent water conservation (approximately 90\% compared to conventional irrigation based vegetable and fish farming \cite{Eck2019}, \cite{plantgrowth2014}) by circulating the same water through both vegetable and fish farming.  
However, a regular manual intervention in the monitoring and control of aquaponics system becomes a huge threat to such an estimated growth.  

In an aquaponics system, primary raw waste is generated by fish and this waste is ammonia rich, which when converted to nitrogen becomes nutritious for plants. 
Therefore, we need to maintain an efficient biological process to convert ammonia to nitrogen in the water, and this process is mainly performed by two types of bacteria: Nitrosomonas bacteria and Nitrobacter bacteria. 
In general, millions of types of bacteria are all around us round the clock in the environment, helping the survival of almost all types of species via some biological process. 
In aquaponics system, beneficial bacteria is another type of bacteria that prevents several types of harmful bacteria to takes its ground in the water and helps in the fish digestion system (particularly within their intestine), and aids in growing healthy root tissue in the vegetables. 
There are several other types of microbes in the water, like fungi, algae, zooplankton, phytoplankton, protozoa, and nematodes, that help in the overall sustainability of the aquaponics and grow automatically provided a right environment is maintained in such a closed water system.
In this paper, we report our experience in building up a Cyber Physical System (CPS) testbed involving a commercial grade aquaponics system.  

A cyber physical system (CPS) is built around a physical process for efficient monitoring and automatic control of the process in general. 
A CPS integrates a set of sensors that sense such a physical process in different capacities for monitoring, a set of actuators that alters different parameters of the process, a set of communicating devices (like programming logic controllers (PLCs)) that locally connects a set of sensors and a set of actuators and communicates with other communicating devices, and a set of devices that provides a human interface to the physical process being monitored and controlled by the CPS (like SCADA or HMI). 
Any physical process that works continuously with a well-defined feedback loop can take advantage of a CPS for its operational efficiency, increased business outcome, enhanced user experience, and avoiding unexpected downtime. 
In this work, we apply tools and techniques of a CPS into an aquaponics system in order to achieve operational efficiency, improved farming yields, safety and security through real time monitoring of the system. 


We have designed and developed a testbed of an aquaponics system to address the problem of continuous monitoring and controlling the major water quality parameters, like pH, DO, TDS, water temperature, air temperature and humidity, that can significantly affect the growth rate of plants (to an extent of dead plants) 
and fish (to an extent of dead fish).
We call the testbed as \emph{CyPhA} (Cyber Physical Aquaponics) system, which is a scaled-down version of a real commercial aquaponics system. 
CyPhA is broadly decomposed into five stages based on the physical processes carried out in an aquaponics system. 
Stage-1 contains the vegetable farming unit of the aquaponics system. 
Note that this vegetable farming does not require any soil and depends only on nutrient rich water; often such farming is called \emph{hydroponics}.
In CyPhA system, this farming unit is designed as a vertical farming system with five vertical levels, where each level can house 40 units and a total of 200 units of seedling of leafy vegetables.  
Stage-2 contains the fish farming unit of the aquaponics system. In our testbed, we consider only one fish water tank having capacity of 80 litres to demonstrate the working of CyPhA system. 
Primary waste is generated in this tank, and this waste water in turn (via Stage-3 and Stage-4) becomes organic fertilizer for vegetable farming. 
Stage-3 performs natural nitrification process and this stage is the first process that helps in turning ammonia rich fish-waste-water into nitrate rich water. 
One of the primary water quality parameters in this water tank Stage-3 is water temperature that helps to build up and maintain bacteria colony for nitrification. 
Stage-4 accomplishes natural bio-filtration process. With the help of bio-balls and bio-filters, this stage further reduces the level of ammonia in the water. 
Some of the primary water quality parameters in this stage are TDS and water temperature. 
The water tanks in these two stages are kept in a relatively dark place where sunlight cannot reach.
Stage-5 contains another water tank that accumulates water either from Stage-4 or Stage-1. 
This stage helps in a regulated release of water into the vegetable farming unit. 

Each stage is monitored and controlled by a Raspberry Pi based edge device, where the edge device is responsible for collecting data directly from sensors in that stage and send it to CyPhA Controller via a CyPhA Gateway. 
It is also responsible for receiving the control commands for the actuators in that stage from CyPhA Controller through a CyPhA Gateway. 
The edge device in each of the five stages communicates with CyPhA Gateway over Wi-Fi network as they both can be within a close proximity.
CyPhA Gateway and CyPhA controller can be located far apart and communicate over public Internet. 
Application data is transferred over MQTT protocol among the devices.

We briefly summarize our contributions as follows:

\begin{itemize}
\item Designed, developed and deployed a CPS testbed, called Cyber Physical Aquaponics (CyPhA) system, for academic research and training in aquaponics systems. The testbed is the first of its kind that involves a complete ecosystem for sustainable food production for modern urbanization. 
\item We have built up a complete solution for continuous monitoring of water and air quality parameter in an aquaponics system that suffer from significant manual intervention. Data visualization dashboard and user command input are facilitated by an open sourced tool called ThingsBoard that can run in a remote system for visual monitoring and control of the complete system.
\item We have designed and implemented a controller, called \emph{CyPhA Controller}, based on an extended finite state machine model that can currently generate control signals for the actuators in Stage-2 of CyPhA system.
\item We have successfully operated our CyPhA system for last 75 days and our analysis shows that the quality of the water, i.e., pH, DO and TDS, that circulates through vegetable and fish farming unit of CyPhA system by applying control commands for water and aeration pumps efficiently. 
\end{itemize}

The rest of the paper is organized as follows. 
Section \ref{sec:relworks} outlines the related works and positions our work appropriately.
Section \ref{sec:systemdesign} provides the details of system design and Section \ref{sec:systembuilt} the implementation of the CyPhA system. 
Section \ref{sec:experimentsetup} provides the details of the experimental setup and data collection for the purpose of the work in this paper. 
Section \ref{sec:results} discusses the results of our analysis and the limitation of the system. 
Finally, we conclude this work in Section \ref{sec:conclusion}.

\section{Related works}
\label{sec:relworks}
Some works have been carried out to reduce manual interventions, along with maximizing the yield and decreasing the use of water and land in plant and fish farming.
Espinosa Faller, et al., simply focus on monitoring an aquaculture recirculating system using a ZigBee-enabled wireless sensor network ~\cite{faller2012journal}. Dan Wang et al., aim to provide users a mobile terminal to monitor and control the household aquaponics system consisting of ornamental fish and hydroponic plants remotely~\cite{Wang2015_07}. TY Kyaw, et al., aim to rectify system abnormality without human interference using collected data and user preset values, but lacks physical dynamics of the system and does not take parameters like DO and nitrite concentration into consideration~\cite{KYAW2017342}. 

Guandong Gao, et al., aim to ensure health of farmed fishes in aquaculture by analysing, predicting, tracking and querying of fishpond water quality data and allowing remote manual equipment control.~\cite{GAO2019105013}. Akhil Nichani et al., also focus on monitoring and allowing end user to make changes in real time on dashboard for aquaponics, but lacks monitoring parameters such as nitrate, nitrite, ammonia and dissolved oxygen and does not provide automated actuation~\cite{2018Akhilintconf}. 
Z. J. Ong et al., focus on reducing electricity cost by exploiting the use of grow lights to boost plant growth rate and allows users to activate and deactivate actuators (fish feeder, water heater and grow lights) but only couple of fish water quality parameters (water temperature and pH) are considered and system is barely adequate for performing analysis~\cite{2019_ICMA}.
Wei Wang, et al., completely focus on providing project-based teaching method which is restricted to the usage of few data and auxiliary variables (water temperature, dissolved oxygen, pH, and conductivity) which are not sufficient for carrying out deep research~\cite{ieeeacess2020}.

AR Yanes et al., enlist essential water and environment monitoring parameters sensed in aquaponics which include nitrification, pH, water temperature, relative humidity, TDS, alkalinity and dissolved oxygen and emphasize dealing with closed loop feedback system that is required to push aquaponics towards smart technologies~\cite{YANES2020121571}. MF Taha et al., provide emphasis on designing a control system flexible enough to allow the monitoring and control of a diverse set of sensors and actuators.~\cite{chemosensors10080303}.

Our proposed system considers aquaponics system as a CPS due to its natural feedback system in the water circulation. 
While designing the system, we have carefully considered the dynamics of the physical process involved in a real commercial grade aquaponics system, for example nitrification, biofiltration and growth of bacteria colony. 
We believe that our proposed system is first of its kind and has a potential to open up a new research area in controlling and monitoring a commercial grade aquaponics system.

\section{System Design}
\label{sec:systemdesign}
\begin{figure}
\center{\includegraphics[width=0.8\linewidth]{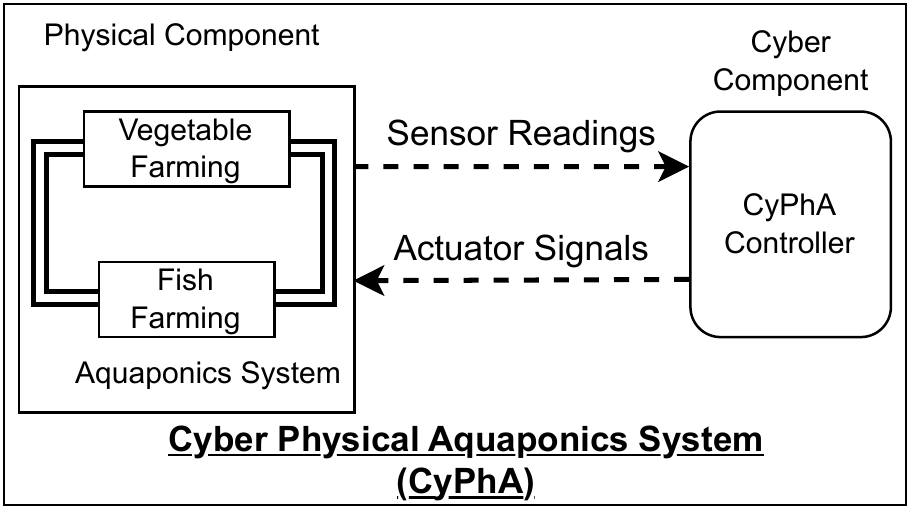}}
\caption{Top level architecture of CyPhA.}
\label{fig:level1cypha}
\end{figure}

Like any other system, the design of our proposed system, CyPhA, is driven by requirements that can be viewed at different levels. 
At the top level, the requirements in CyPhA can be written as follows:
\begin{itemize}
    \item \emph{Continuously} monitor and control water flow through vegetable and fish farming unit in the aquaponics system.
\end{itemize}
Based on this requirement at the top level, the high level architecture of our proposed CyPhA is shown in Figure \ref{fig:level1cypha}. 
The physical component in CyPhA consists of a traditional aquaponics system. 
The cyber component in CyPhA consists of the monitoring and controlling unit that receives sensor readings and generates actuator signals.
This high level requirement is decomposed into several requirements in a hierarchical fashion based on a decomposition of the physical component in the CyPhA system, discussed in the following subsections.  

\subsection{System Design in the $2^{nd}$ Level}
In the $2^{nd}$ level, we have two broad requirements based on the two separable units of an aquaponics system, i.e., vegetable farming unit and fish farming unit.
\begin{itemize}
    \item \textbf{RQ1:} \emph{Continuously} monitor and control water flow through vegetable farming unit.
    \item \textbf{RQ2:} \emph{Continuously} monitor and control water circulation through fish farming unit.
\end{itemize}
The vegetable farming unit is responsible for hosting leafy vegetables, which we consider a vertical farming system.
Vegetables in this system do not require any soil. 
A bunch of vegetable saplings is put in a disposable water glass with wood dust and chips so that full-grown plants catch hold to the glass. 
Four to five vertical holes are made in the wall of the plastic glass so that plant roots can come out of the glass and immerse in water flowing through the pipe.
Note that the water flowing through the pipe is nutritious because it is both nitrogen and ammonia rich. 
In the current system, the complete vegetable farming unit is considered as one stage in CyPhA system. 

Actually, the fish farming unit is responsible for hosting fish. But, to overcome ethical issues, we have considered using fish water that we collect from a local fishery farm that farms  Murrel/Sole fishes. 
This ammonia rich effluent water is suitable for vegetable farming after nitrification. 
Compared to non-eatable aquarium fish, farming edible fish generally has commercial implications. 
Fish water that we consider in our fish farming unit contains a higher amount of ammonia due to edible fish being farmed, compared to that of non-eatable aquarium fish. 
A higher concentration of ammonia in the fish tank is life-threatening for the fish, and therefore continuous monitoring of water quality is required to maintain permissible fish water parameters.
Thus, the fish farming unit requires greater attention to monitoring and control, compared to vegetable farming.

\begin{figure}
\begin{subfigure}{.5\linewidth}
\centerline{\includegraphics[width=.45\linewidth]{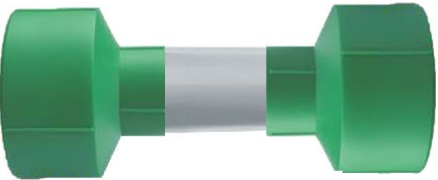}}
\caption{Pipe Reducer and Inter-connector.}
\label{fig:pipeReducer1}
\end{subfigure}
~
\begin{subfigure}{.5\linewidth}
\centerline{\includegraphics[width=.3\linewidth]{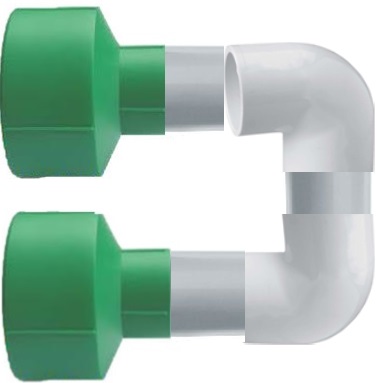}}
\caption{Pipe Reducer and Inter-connector. }
\label{fig:pipeReducerAndConnector1}
\end{subfigure}
\caption{Model of PVC Pipe Connectors. }
\end{figure}

\begin{figure}
\centerline{\includegraphics[width=.75\linewidth]{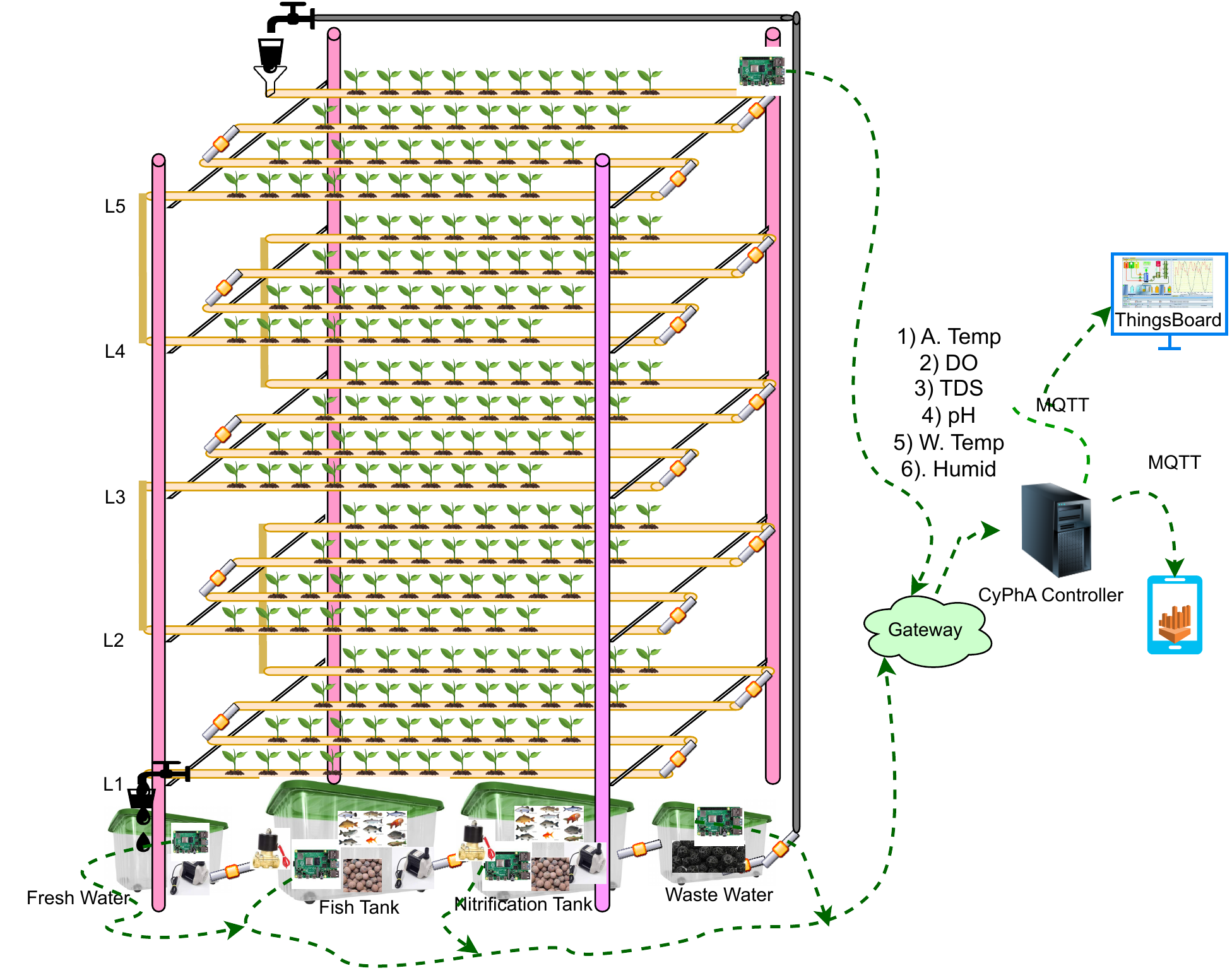}}
\caption{System Model of CyPhA.}
\label{fig:system_model}
\end{figure}

The physical structure of CyPhA system satisfying these two high level requirements is shown in Figure \ref{fig:system_model}. 
In the RQ1, we have designed a vertical farming unit that has five vertical levels, L1 (the bottom level) to L5 (the top level).  
One water pipe having 3 inches of diameter is used for flowing effluent water. 
The top side of this pipe has 10 holes, each of 2 inches in diameter, and one such hole can hold one unit of vegetable saplings, i.e., one glass of saplings. 
Thus, one pipe can hole 10 vegetable units and every level has four such pipes. 
The end points of these pipes are connected in a cascading manner using 3 to 1 inch reducer adapter pipes (as shown in Figure \ref{fig:pipeReducer1} and  Figure \ref{fig:pipeReducerAndConnector1}). 

Such an arrangement allows excess water to remain stagnant in the pipe where the vegetable root is immersed in water, and such water becomes a breeding ground for useful bacteria. 
Further, the one end of the last pipe in a level is connected to the one end of the last pipe in the next level. 
Thus, an overall stretch of the pipes becomes about $5 \times 4 \times 5 = 100$ feet through which effluent water flow, and a total of $10 \times 4 \times 5 = 200$ units of vegetable saplings are hosted in this stretch of water pipe.
So, the overall requirement in the vegetable farming unit is to ensure that water is flowing through the pipes at a regulated rate. 
In the current version of the work, we do not decompose any further the vegetable farming unit. 
After the circulation, water is collected in a storage tank, from which it is supplied to the fish farming subsystem.
The vegetable farming unit along with the storage tank is considered Stage-1 in the CyPhA system. 

\subsection{$3^{rd}$ Level Design of Fish Farming Subsystem}
The requirements in the fish farming unit are decomposed into three more detailed requirements. 
\begin{itemize}
    \item \textbf{RQ2.1:} \emph{Continuously} monitor and control the amount and the quality of water in the fish tank.
    \item \textbf{RQ2.2:} \emph{Periodically} transfer the effluent water from the fish tank into the natural nitrification water tank and monitor the quality of water in the nitrification tank.
    \item \textbf{RQ2.3:} \emph{Periodically} transfer the water into a biofiltration water tank and monitor the quality of water in the biofiltration tank.
    \item \textbf{RQ2.4:} \emph{Continuously} supply effluent water into vegetable farming unit. 
\end{itemize}
RQ2.1 indicates that the quality of the water needs to be monitored using appropriate sensors and if the quality degrades, then water needs to be shifted from the fish tank to the nitrification storage tank using pumps. 
Once water shifting is done, fresh water needs to be brought into the fish tank, therefore the controller of Stage-1 needs to be notified. 
RQ2.1 forms Stage-2 in the CyPhA system. 

\begin{figure}
\begin{subfigure}{.23\linewidth}
\centerline{\includegraphics[width=.8\linewidth]{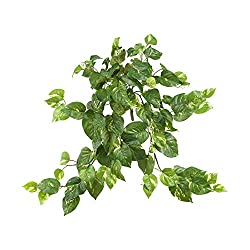}}
\caption{Pothos}
\label{fig:pothos_water_plant}
\end{subfigure}
~
\begin{subfigure}{.23\linewidth}
\centerline{\includegraphics[width=.8\linewidth]{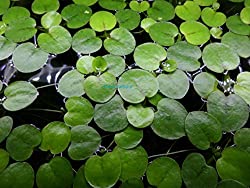}}
\caption{Frogbit. }
\label{fig:frogbit_water_plant}
\end{subfigure}
~~~
\begin{subfigure}{.23\linewidth}
\centerline{\includegraphics[width=.7\linewidth]{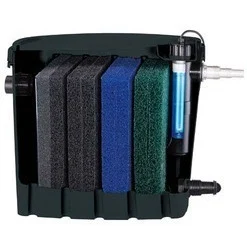}}
\caption{Biofilter }
\label{fig:biofilter}
\end{subfigure}
~
\begin{subfigure}{.23\linewidth}
\centerline{\includegraphics[width=.9\linewidth]{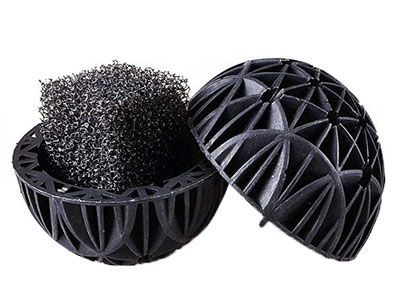}}
\caption{Bioballs}
\label{fig:bioballs}
\end{subfigure}
\caption{Natural and artificial conversion of ammonia to nitrate using water plants and bio-balls \& bio-filter.}
\end{figure}

RQ2.2 indicates that ammonia rich effluent water undergoes a natural process of nitrification. 
Several water plants, like water lily and water lettuce, pothos, and Amazon frogbit as shown in Figure~\ref{fig:pothos_water_plant} and Figure~\ref{fig:frogbit_water_plant}, help in the natural nitrification process.

Moreover, such plants create a healthy breeding ground for bacteria that helps in nitrification, like Nitrosomonas, Nitrosococcus, Nitrosolobus, Nitrosovibro, Nitrobacter, Nitrospira, and Nitrospina. 
It is important to maintain appropriate water quality parameters like stable pH, DO, temperature, and no UV light, so that the bacteria colony survives via reproduction. 
For this work, we do not pay great attention to this requirement, but it is an unavoidable process in the aquaponics system. 
RQ2.2 forms Stage-3 in the CyPhA system. 

RQ2.3 indicates the requirement of a biofiltration process in the aquaponics system. 
The biofiltration process helps in converting ammonia and nitrite into nitrate, which is facilitated by beneficial bacteria. 
Basically, biofilters as shown in Figure~\ref{fig:biofilter}, block any remaining solid fish waste from flowing into the vegetable farming unit, i.e., Stage-1. 
Also, the subsystem is required to allow beneficial bacteria to build up and sustain it colony. 
Often, the circulation of bio-balls using air pumps in the water is performed in order to create a healthy environment for the bacteria colony. 


Figure~\ref{fig:bioballs} depicts Bio-balls, which  are used to increase the surface area within a small tank, and the increased surface is good for bacterial growth. 
In this initial setup, we do not focus much on this requirement, but this is an essential part of an aquaponics system. 
RQ2.3 forms Stage-4 in the CyPhA system. 

RQ2.4 indicates that water needs to be supplied at a slow rate but for a long duration into the vegetable farming unit, i.e., Stage-1. 
Basically, it indicates that we need to install a storage tank for effluent water, from which water will be supplied to Stage-1. 
We have incorporated a mechanism based on gravitation to produce a slow rate of water supply using reduced diameter water pipe. 
In general, the transfer of water from one tank in one stage to another tank in the next stage requires time in the range of [5, 10] minutes, whereas, we need to supply water for a longer duration, in the range of [2, 4] hours. 
Therefore, a water pipe of diameter 0.5 to 1.0 inch is used for the transfer of water from one tank to another. 
But, for the supply of water into Stage-1, we use a pipe of diameter 0.25 inch. 
RQ2.4 forms Stage-5 in the CyPhA system.

\subsection{Design of Cyber System}
\begin{figure}[htbp]
\centerline{\includegraphics[width=0.9\linewidth]{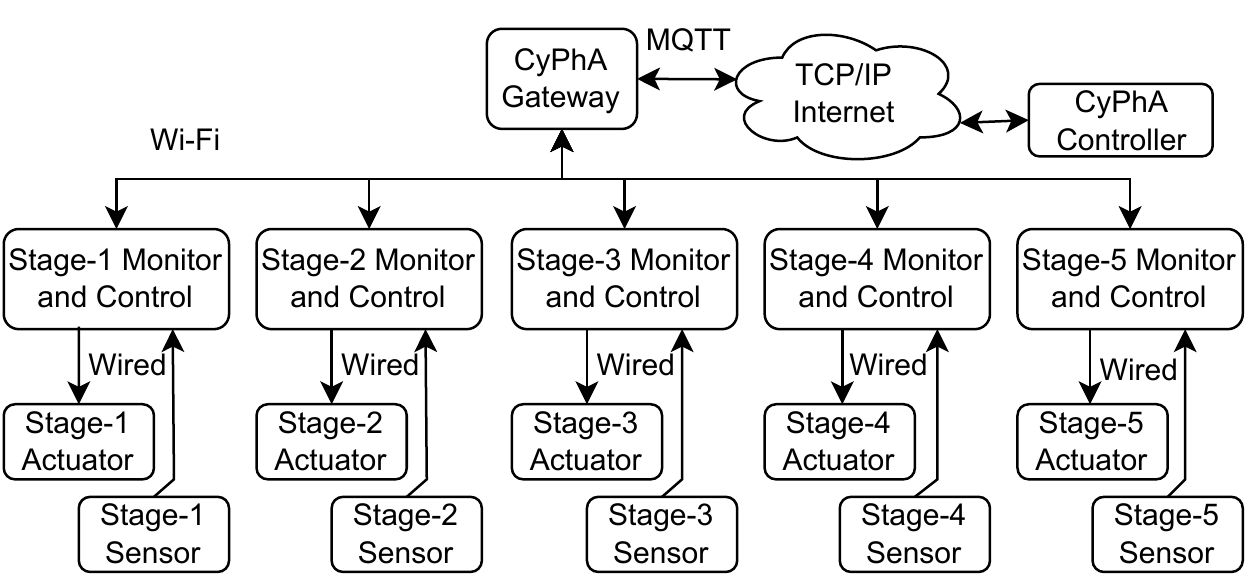}}
\caption{Architecture of Communication Network in CyPhA System.}
\label{fig:comm_arch_CyPhA}
\end{figure}

The communication among the sensors, the actuators and the CyPhA Controller (Figure \ref{fig:arch_CyPhA}) is decomposed into subcomponents based on the physical structure of our CyPhA system. 
The design of the networking part of the cyber system in CyPhA plays an important role in the load on the communication network and energy consumption due to transmission. 
Figure \ref{fig:comm_arch_CyPhA} shows the design of the communication network used in the CyPhA system. 

At the lowest level, the sensors and the actuators in a stage are connected over a wired network to the controller in the stage itself. 
We divide the sensors and actuators into two parts because sensing and actuating frequency is different and are handled by two different units in the controller of the respective stage. 
For instance, the monitor and controller in Stage-1 is isolated from that in Stage-2. 
If there is a requirement of communication across stages, then the communication is routed all the way through the CyPhA Controller. 
Because the CyPhA controller and the CyPhA physical component need not be located in close proximity, we consider TCP/IP Internet to facilitate the communication. 
However, the stage-wise controllers in the five stages communicate to a local CyPhA gateway, which in turn communicates with the CyPhA Controller. Such an arrangement helps in availing public or private network setup for long range communication, allowing a higher physical separation between the CyPhA Physical Component and the CyPhA controller. 
For instance, the gateway can be equipped with LoRAWAN communication infrastructure, that provides a larger communication range compared to Wi-Fi network. 
Further, the gateway can be helpful for activating security and safety mechanisms for accessing/manipulating the local CyPhA physical infrastructure and enable effective cyber-security mechanisms for communication with the CyPhA controller. 
We used MQTT as an application layer protocol for carrying sensor data or actuator signal between the CyPhA Gateway and the CyPhA Controller. 

Among other, we found MQTT as most suitable for the design of communication infrastructure due to its publish-subscribe architecture of communication. 
The stage-wise monitor and controller play roles of both publisher and subscriber. 
Acting as publisher, it sends sensor data from its stage to the CyPhA controller and acting as subscriber, it receives actuator signals for the actuators within its stage. 
A Human Machine Interface (HMI) can be integrated in CyPhA system that can act as both subscriber and publisher. 
Acting as subscriber, it periodically receives both sensor readings and actuator status for effective visualization and acting as publisher, it can receive manual commands for actuators from the users and then send it to the controller, which in turn published the control signals appropriately to the respective stage(s). 
Thus, our communication network highly depends on the reliability of MQTT-based communication over TCP/IP Internet and on the Wi-Fi network for local communication.

\subsection{System Design Supporting Safety Requirements in CyPhA}
Like any other CPS system, safety requirements in addition to functional requirements builds a higher operational confidence in CyPhA system. 
Safety requirements can be broadly categories into two types. 
First, safety measures during handling of physical components, e.g., wearing gloves for touching normal water or fish water or plants, to avoid any contamination or disease spread. 
While we follow certain safety measures for this, we do not consider this type of safety measure a non-technical in nature and hence our scope of this paper. 
Second, operational safety measures that specify guidelines for the maintenance of water quality and water circulation for both fish and vegetable farming.
This type of safety requirement can be satisfied by the monitoring and control of an aquaponics system. 
Some of the requirements in this category are discussed in the following, along with the design considerations of a CyPhA system that satisfies the requirements. 

\begin{itemize}

\item Water Circulation: Regular water circulation through both the vegetable and fish farming units in the aquaponics system helps to reduce the risk of growing unwanted or harmful bacteria, like Sulphate-Reducing Bacteria and De-Nitrifying Bacteria (collectively called pathogenic bacteria), that can be life-threatening for fish or destroying the vegetables. 
A commonly cited guideline \cite{gogreen} for a densely-stocked fish tank is to circulate the water 2 times/hr. For example, if a fish tank is of 1000 litres, the water flow rate should be 2000 litres/hr, so that every hour the water is circulated twice. 
On the other hand, at low stocking densities this turnover rate is not mandatory, only 1 time/hr is sufficient. 

Consider such a requirement as part of the design of the controller in every stage. 
A time constraint is applied to activate the water circulation process in the controller. 
Because such circulation of water has to happen in every stage, the appropriate control signals are sent from the CyPhA Controller to all the stage in a coordinated manner to avoid water overflow in any particular stage.

\item Aeration: Aeration is a process of mixing oxygen in water so that the level of dissolved oxygen (DO) can be increased in order to help healthy survival of fish, plants and bacteria. 
Though a different level of DO is required in different species of water animals, 4-5mg/litre of DO is a typical requirement for medium density of fish tank \cite{somerville_safety_2014}. 
Appropriate level of DO is required also for natural decomposing any waste in water.

To address this requirement, CyPhA system considers monitoring DO level in each of the five stages. 
As an initial implementation, we consider air stone with air pump for aeration in the water tanks in each of the stages. 
A typical specification of such aeration pump operates with 220-240v and 50/60/Hz frequency having 3L of air per minutes as the output capacity. 
A control of this type aeration pump in a stage is implemented in a stage-wise controller. 
Activation or deactivation of this pump depends on the current level of DO in a water tank in the stage. 


\item Water Temperature: 
In the aquaponics system, both air and water temperature are equally important. Though there is an inherent one directional dependence where, air temperature can have direct impact on water temperature, continuous monitoring of air temperature and controlling proper ventilation in the surrounding environment is part of safety requirement. 
Water temperature has a direct impact on both DO and toxicity due to the ionization of ammonia. 
A higher water temperature tends to have less DO and more unionized ammonia. 
At higher air temperature, plants suffer from low absorption of calcium. 
Certain studies ~\cite{somerville_safety_2014} have shown that a proper combination of fish and plant in the aquaponics system could reduce the risk due to variation in air and water temperature. 

In CyPhA system, we satisfy this requirement selecting an appropriate combination of fish water and plant in the system and by consider that the complete system be carefully exposed to open air. 
Additionally, we plan to consider ventilation fans to satisfy this requirement.

\item pH level: Plants prefers slightly acidic water with PH ranges from 6 to 6.5 and fish prefers alkaline water with range from 7 to 8.5. 
According to some commercial firms ~\cite{gogreen}, an ideal range to maintain a good balance of PH for an effective growth of both plants and fish is [6.8, 7.2].  Additionally, some research works ~\cite{journal_amri_2022} suggest that ideal pH for safe operation of aquaponics system [6, 8.5]. 
Similarly, ideal TDS range is [200, 400] PPM, as suggested by some commercial firms ~\cite{james2019_naver}. 

In CyPhA system, we satisfy this requirement on pH or TDS by continuously monitoring and controlling the circulation of water. 
If the pH level in water in the vegetable farming unit is not within range, then we circulate water within Stage-1 by directly providing water from Stage-1 to Stage-5. 
Such a case is logged for future reference. 
After a number of such re-circulations, if pH or TDS is not improved, we consider that water is no more suitable for CyPhA system.
In such extreme case, the system rejects the contaminated water and signal for intake of fresh water from external sources for the aquaponics. 

\end{itemize}

\section{Implementation of CyPhA System}
\label{sec:systembuilt} 
In this section, we mainly discuss (in a bottom-up approach) the implementation of the cyber component of CyPhA system. 
\subsection{Implementation of Stage-2 Cyber Component} 
\begin{figure}[htbp]
\centerline{\includegraphics[width=0.7\linewidth]{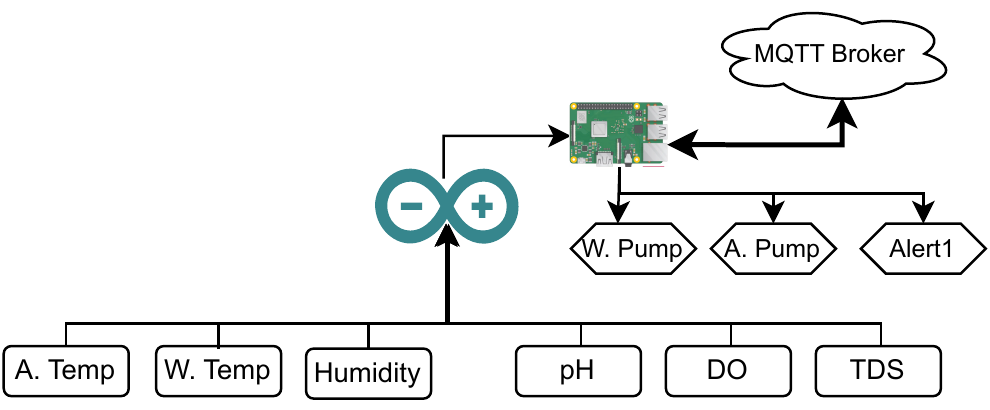}}
\caption{Stage-2 Sensing and Actuating Model.}
\label{fig:arch_CyPhA}
\end{figure}

The schematic view of the connection of the sensors and the actuators in a particular stage (Stage-2) is shown in Figure \ref{fig:arch_CyPhA}. 
In order to monitor the quality of water in the fish tank in Stage-2, we have considered Arduino UNO compatible sensors for pH, DO, TDS, water temperature, air temperature and humidity. 
Specifically, each of the water quality sensors (i.e., pH, DO, TDS) is exclusively connected to one of six analog pins in the microcontroller, whereas the other three sensors (water \& air temperature and humidity) are connected to the digital pins of the controller.  
Every two seconds, we take the readings from each of these sensors and send it to the edge device (i.e., a Raspberry Pi) in Stage-2. 

The actuators in the fish tank are connected to a Raspberry Pi Model 3 microcomputer. 
Specifically, each of the water pump, aeration pump and LED alert indicators is wire connected using a digital GPIO pin in the microcomputer. 
As an initial setup, the microcontroller Arduino UNO is connected to the microcomputer Raspberry Pi over USB interface and hence the communication between the two is over the serial port in the microcomputer. 
Thus, the microcomputer collects sensor data whenever some data is available over the serial interface, and it acts as an edge device for Stage-2, that communicates with the CyPhA Controller via CyPhA Gateway.

CyPhA Gateway can be implemented in one of two ways. Either it is a separate microcomputer or an edge device in one of five stages. 
For the purpose of this paper, we consider the microcomputer in Stage-2 is acting as the CyPhA gateway. 
Basically, the gateway facilitates the communication between the physical component of CyPhA and the CyPhA controller, assuming that the two will be separated geographically beyond the range (maximum of 100 m) of a typical Wi-Fi network. 
Additionally, the gateway is responsible for cyber or network security mechanisms in order to provide overall safety and security of the CyPhA system. 
For the purpose of this paper, we have incorporated certain mechanism for providing data integrity service to the CyPhA related being transferred between the gateway and the CyPhA Controller.

\subsection{Implementation of CyPhA Controller and its Cyber Component}
The functionalities of CyPhA controller functionality is broadly categorized into three main tasks:
\begin{itemize}
    \item \textbf{Task 1:} Receive the sensor readings from the edge device of the CyPhA system.
    \item \textbf{Task 2:} Process the sensor readings and generate appropriate control signal based on an extended finite state automata.
    \item \textbf{Task 3:} Publish the received sensor values and processed actuator signals to the data-analytic server and edge device, respectively.
\end{itemize}

\begin{figure}[h]
 \centering
\begin{center}

\begin{tikzpicture}[shorten >=1pt,node distance=3cm,auto]
  \tikzstyle{every state}=[fill={rgb:black,1;white,10}]

  \node[state, initial] (q_1)  {$q_1/00$};
  \node[state]           (q_2) [below of=q_1]     {$q_2/01$};
  \node[state] (q_3) [ right of=q_1] {$q_3/10$};
  \node[state]           (q_4) [below of=q_3]     {$q_4/11$};

  \path[->]
  
  (q_1) edge [loop above]  node {a} (   )
        edge [bend left]  node {b} (q_2)
        edge [bend left]  node {c} (q_3)
        edge [bend left=10] node {d} (q_4)
        
  (q_2) edge [loop left]  node {b} (   )
        edge [bend left]  node {a} (q_1)
        edge [bend left=10]  node {c} (q_3)
        edge [bend left] node {d} (q_4)

  (q_3) edge [loop above] node {c} (   )
        edge [bend left]  node {d} (q_4)
        edge [bend left]  node {a} (q_1)
        edge [bend left=10] node {b} (q_2)

  (q_4) edge [loop right] node {d} (   )
        edge [bend left]  node {c} (q_3)
        edge [bend left]  node {b} (q_2)
        edge [bend left=10] node {a} (q_1);
\end{tikzpicture}

\end{center}
\caption{Extended Deterministic Finite State Machine that controls Stage-2 of CyPhA system. Transitions: \{a: pH\&DO in safe range, b: pH in safe range, DO is not in safe range, likewise other transitions\}. Output: \{00, both WP (water pump)\&AP (aeration pump) are OFF; 01: WP is OFF and AP is ON; likewise other output states \} }
\label{fig:senandact}
\end{figure}
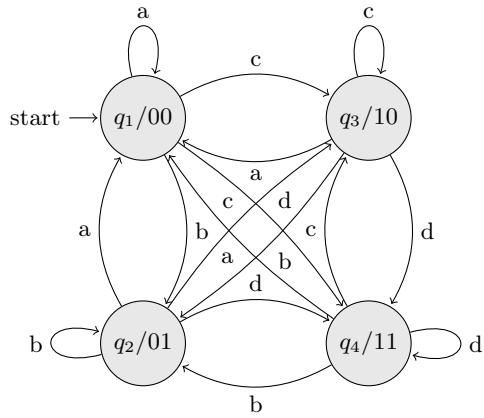

We have implemented these functionalities in three broad submodules within the controller. 
The controller is coded in Python and associated packages. 

\textbf{Task 1:} The messages are received from the edge devices in the form of JSON objects. 
A JSON parser extracts the values from the received JSON string. 
The extracted values are then sent to another module for integrity check.
Once successful, the sensor values are passed to the next module.

\textbf{Task 2:} The verified sensor values are supplied as input to an extended finite state machine (shown in Figure \ref{fig:senandact}) that generates control signals as output. 
Note that the state machine shown in Figure \ref{fig:senandact} is only to control Stage-2 of the CyPhA system. 
In our current implementation, there are two actuators, a water pump and an aeration pump. 
Whenever the state machine is in a particular state, the output is fixed, similar to a Moore machine. 
For instance, in $q_1$ state, the output is $00$ indicating the fact that both the pumps be switched off. 
The output generated from the state machine are sent back as control signals to the edge device over MQTT. 

The state $q_1$ of the state machine is the start state, where both the water pump and aeration pump are in OFF state.
This state indicates that both pH and DO are in \emph{safe} range. As long as the sensor readings of pH and DO are in this range, the output is 00; this is indicated by the transition denoted by "a". The transition denoted by "c" is corresponding to the fact that the received pH reading is not in \emph{safe} range and the received DO reading is in \emph{safe} range. 
The machine transits to state $q_2$ and the output is 10 indicating the water pump should be switched on, and the aeration pump be switched off.
The rest of the transitions are similar. 
The safe range of each of the sensor are provided via a configuration file so that the state transitions can be controlled without disturbing the source code much.  


\noindent\textbf{Task 3:} The controller has a data publishing module that sends the control signals generated by the state machine. The data publishing module is also responsible for data logging (both sensor data and actuator signals) at regular intervals of time.
\begin{figure}[h]
\centerline{\includegraphics[width=.9\linewidth]{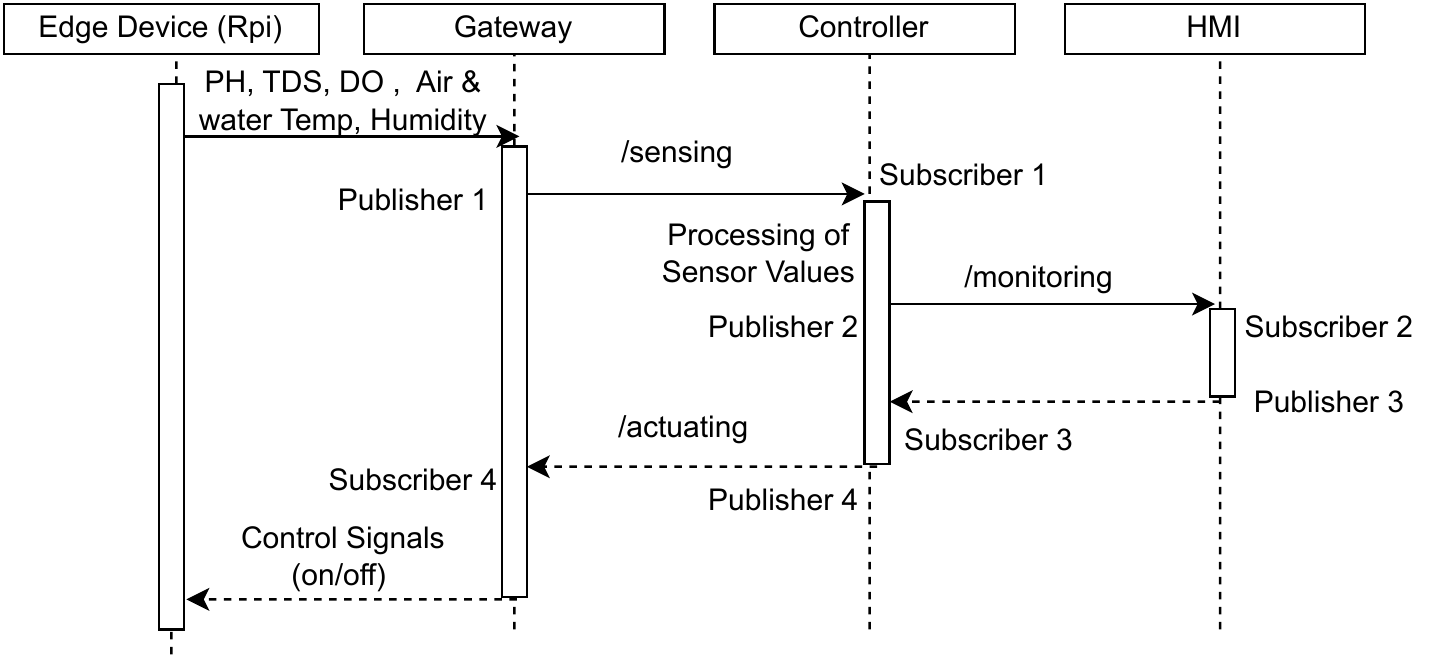}}
\caption{CyPhA Sequence Diagram.}
\label{fig:CyPhA_activity_diagram}
\end{figure}

The cyber component of the controller implements MQTT (Message Queue Telemetry) protocol in the application layer. 
Therefore, the controller receives the sensor reading by subscribing to a topic names as "Stage2Sensing". 
Edge device in Stage-2 is responsible for publishing the sensor values on this topic. 
The actuator signals are published by the controller with a topic named as "Stage2Actuating", which is received by Stage-2 edge device. 
Finally, the controller also subscribes to a topic called as "Stage2ManualActuating", on that an HMI module publishes control signals. 
Manual control signal are given a higher priority over controller generated control signals. 
A sequence diagram is shown in Figure~\ref{fig:CyPhA_activity_diagram} to describe a sequence of high level activities undertaken by the controller when a sensor reading is received.


\subsection{Implementation of Physical Component of CyPhA system}
The physical component of the CyPhA system comprises mainly with suitable hardware that facilitates vegetable and fish farming in loop.
Figure~\ref{fig:exp_setup} shows the overall implementation of the physical component in CyPhA system. 
The setup is created in a laboratory environment within an area of about 80 sq.ft.

Stage-1 of CyPhA system is essentially a vertical vegetable farming unit with five levels. 
The vertical and horizontal connectivity in the PVC pipes throw which water circulates is shown as a vertical farming garden in Figure \ref{fig:exp_setup}. 
Stage-2 is a fish farming unit consisting of 80 litres of water tank. 
Stage-3 is hosted in another 80 litres of water tank that contain one unit of each of water lily and water lettuce. 
Stage-4 is hosted in another 80 litres of water tank containing 1 kg of pebbles stones that is used for water purification along with bio-balls and bio-filters. 
Stage-5 is hosted in 20 litres of water tank which is periodically filled in around 2 minutes time and drained out in the span of around 2 hours.

\section{Experimental Setup}
\label{sec:experimentsetup}
In the experiments, our primarily focus on Stage-2 in CyPhA system. 
Stage-2 has four sensors and two actuators. 
All the sensors are used in its default calibration, like 1.5V, 2.0V and 3.0V indicate pH values as 4.0, 7.0 and 9.0 respectively. 
The electrode probe in DO sensor is filled with the potassium chloride electrolyte solution and diffuses through the thin membrane to calculate DO in water. 
DHT11 sensor is used to measure air temperature and humidity.
DS18B20 sensor is used to measure water temperature. 
Water pump with capacity 900L/h and aeration pump with capacity of 3L/m operate in 165-220V, so we use relay (SONGLE) to control with 5V supplied by the edge device. 

Stage-2 edge device is implemented using Raspberry Pi (Model 3 B+)  that sends the sensor value to the CyPhA controller via gateway. The CyPhA controller and the HMI for data visualization are configured on two Ubuntu 18.04 based desktops. 
The CyPhA controller is implemented in Python 3.0. 
The Controller considers a configuration file that provides user defined thresholds values of different sensors, e.g., \emph{TDSpermissible} [300, 500], \emph{PHpermissible} [6.5, 8.5], \emph{DOpermissible} [3.5, 5], to generate control signals.

\section{Data Acquisition and Result Analysis}
\label{sec:results}

For the purpose of this paper, we have closely monitored our system for $3\times 24 = 72$ hours and collected the dataset, though the system is running for last 75 days. 
The dataset contains 9 attributes, namely timestamp (at which sensor data is received at the controller), pH, TDS, DO, water temperature, air temperature, humidity, water pump, aeration pump. The attributes are related only Stage-2 of CyPhA system.
A total of about 30,000 records with a size of 2.8 MB is collected. 
The system is running in a contained environment with the least or no manual intervention. 
We have also noticed the life cycle of plants in our system. While life cycle of edible fish can be longer than a year,
typical life cycle of lettuce is about 50 days (from seed to harvest) and it is about 40-45 days for Coriander. 
Therefore, we have planned to observe the running of the system for about 50-60 days and investigated the growth rate. 
As an initial experiment, we have enforced water circulation through both fish farming unit and vegetable farming unit only for one time in 24 hours time.
It has been observed that the plant can survive for a minimum of 15 days without requiring extra water, and the increase in the weight of plants noticed is satisfactory. 

\begin{figure}[t]
\begin{subfigure}{.3\linewidth}
\centerline{\includegraphics[width=\linewidth]{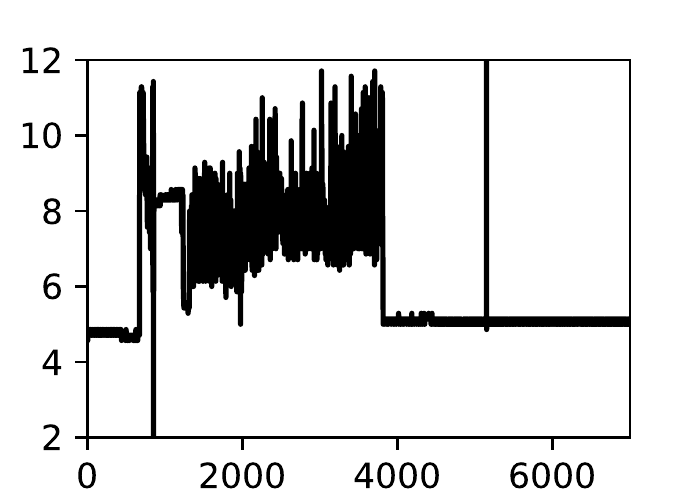}}
\caption{First day: pH.}
\label{fig:ph_first}
\end{subfigure}
~
\begin{subfigure}{.3\linewidth}
\centerline{\includegraphics[width=\linewidth]{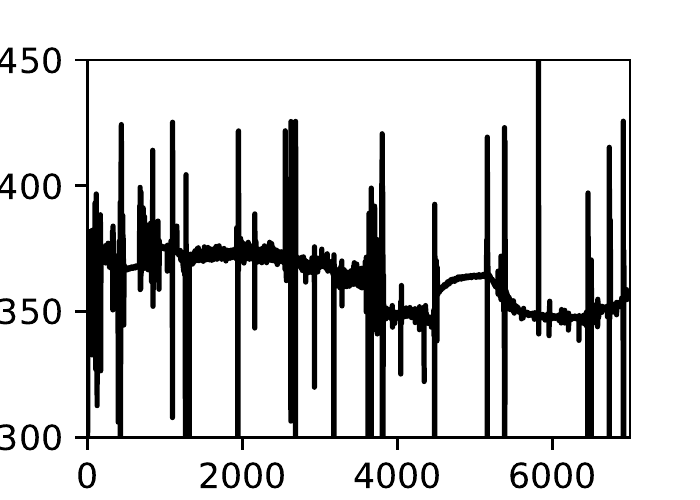}}
\caption{First day: TDS.}
\label{fig}
\end{subfigure}
~
\begin{subfigure}{.3\linewidth}
\centerline{\includegraphics[width=\linewidth]{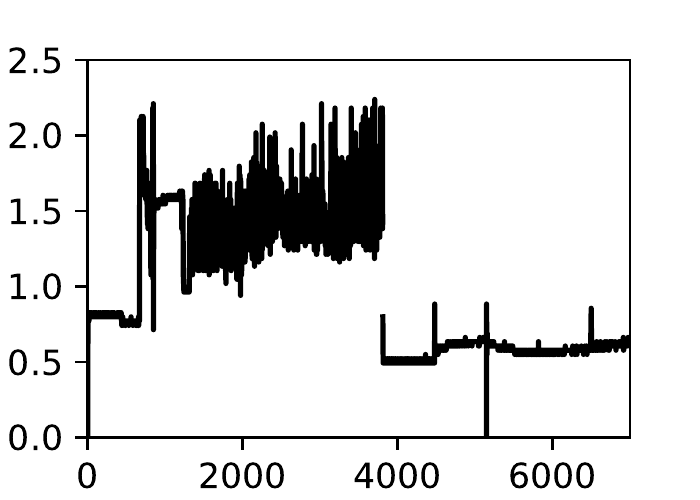}}
\caption{First day: DO.}
\label{fig:do_first}
\end{subfigure}
\begin{subfigure}{.3\linewidth}
\centerline{\includegraphics[width=\linewidth]{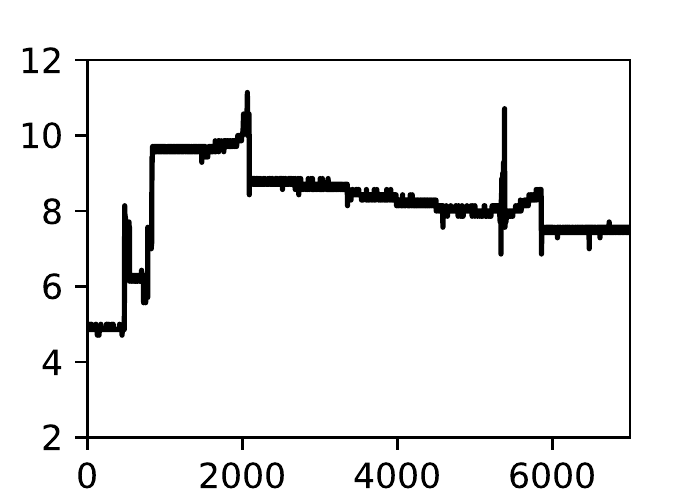}}
\caption{Second day: pH.}
\label{fig}
\end{subfigure}
~
\begin{subfigure}{.3\linewidth}
\centerline{\includegraphics[width=\linewidth]{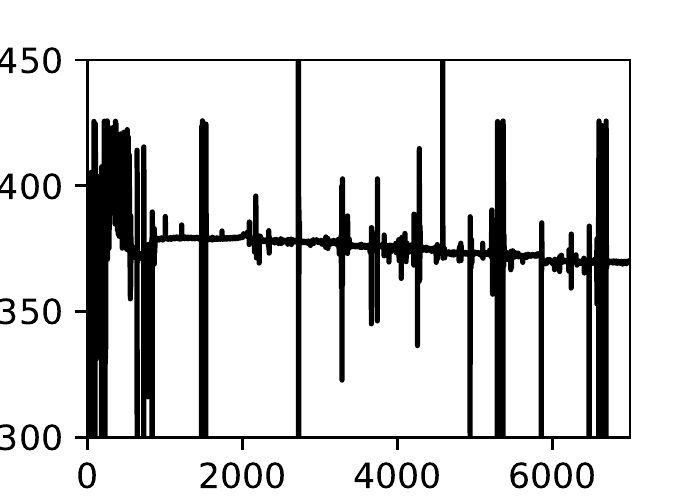}}
\caption{Second day: TDS.}
\label{fig}
\end{subfigure}
~
\begin{subfigure}{.3\linewidth}
\centerline{\includegraphics[width=\linewidth]{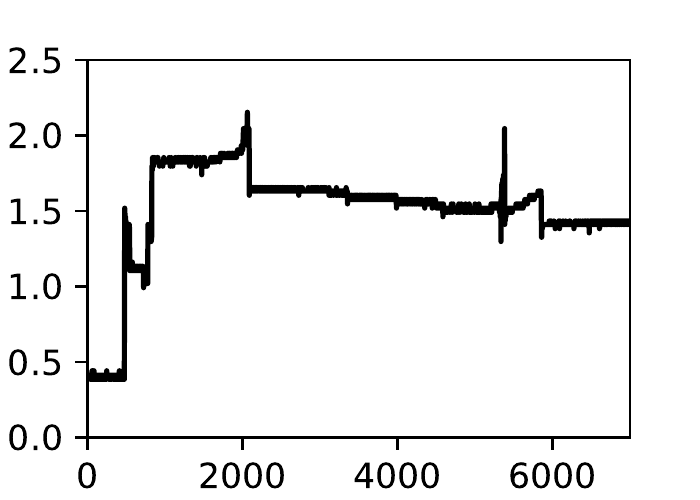}}
\caption{Second day: DO.}
\label{fig}
\end{subfigure}
\begin{subfigure}{.3\linewidth}
\centerline{\includegraphics[width=\linewidth]{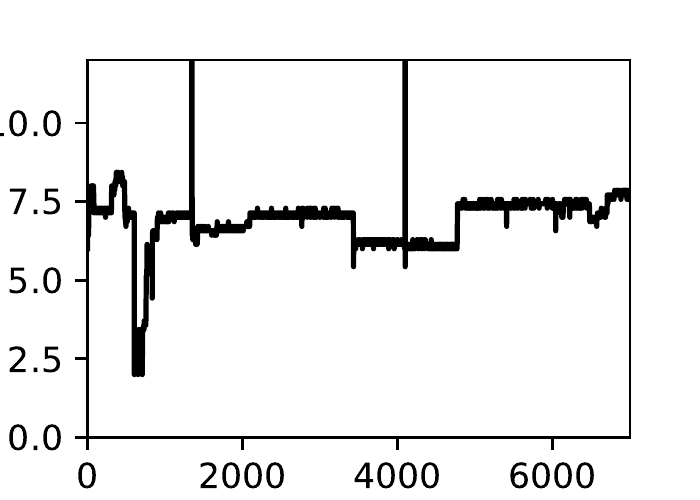}}
\caption{Third day: pH.}
\label{fig:day3pH}
\end{subfigure}
~~~~
\begin{subfigure}{.3\linewidth}
\centerline{\includegraphics[width=\linewidth]{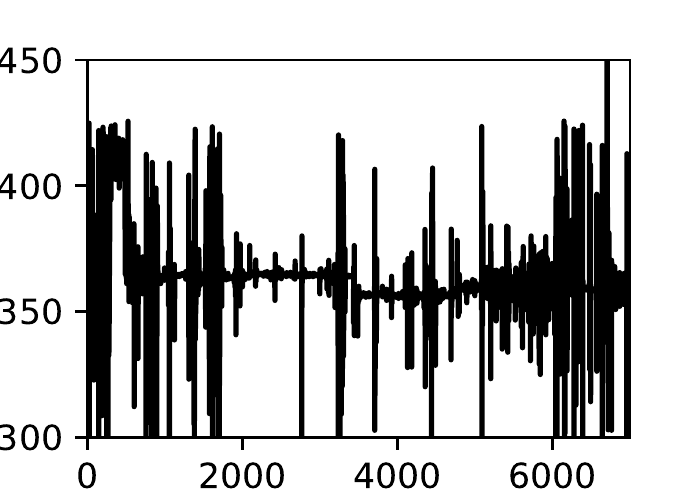}}
\caption{Third day: TDS.}
\label{fig}
\end{subfigure}
~~~~
\begin{subfigure}{.3\linewidth}
\centerline{\includegraphics[width=\linewidth]{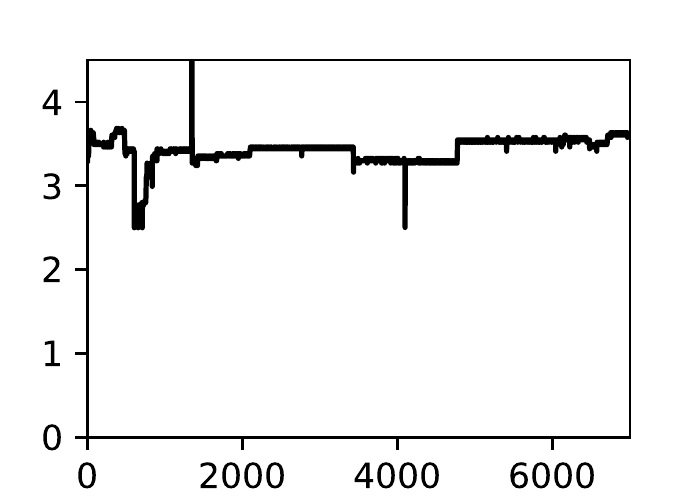}}
\caption{Third day: DO.}
\label{fig:do_thirty}
\end{subfigure}

\caption{Water Parameters for 3 arbitrarily chosen Days in Stage-2.}
\label{fig:water_param}
\end{figure}






\subsection{Analysis of Water Quality in Stage-2}
Figure \ref{fig:water_param} shows the time series data of water quality parameters in arbitrarily chosen three days out of 75 days of our plant operation of CyPhA system. 
The value of pH seen to be highly fluctuating in the initial days (Figure \ref{fig:ph_first}). 
Several factors contributed to this fluctuation, from sensor placement to water transfer to water circulation to fresh plantation. 
Eventually, we were able to normalize pH when water circulation became regular and the plant started growing roots and immersed into water (Figure \ref{fig:day3pH}). 

TDS can be introduced into fresh water by both fish and plant roots. 
In general, a TDS sensor calculates the amount of dissolve solids based on the conductivity of the water; a higher conductivity results in a higher TDS value. 
We assume that the movement of fish water near or far to or from the TDS sensor play a critical role in such fluctuations in the fish tank. 

DO is essential for both maintaining fish health via its food digestion and waste food decomposition within the fish tank. 
Typically, DO below 2 mg/L stops the nitrification process. 
In Stage-2 of the CyPhA system, we were able to maintain DO in the range [0.5,2.5]mg/L (Figure~\ref{fig:do_first}). 
But, we have observed that it goes down to 1 mg/L or below frequently.
With no further delay after this observation, we have installed an aeration pump in this tank, 
and the level of DO rise upto 3.5 mg/L (Figure~\ref{fig:do_thirty}).

\subsection{Challenges in Aquaponic System}
Building an operational aquaponics system as a CPS requires expertise in the fields of biology, electrical, civil and computer science and engineering.
In biology, it requires knowledge in water cycle, nitrogen cycle, food chain in water animals and vegetables, and hydroponics and aquaculture species. 
Maintaining the right values of different water parameters driving both the farming is challenging. 
For example, pH stabilization is an important process for all living organism to survive and grow in a contained environment. 
The knowledge of natural nitrification through autotrophic bacteria, namely nitrosomonas, and providing the right environment to maintain their colony is another challenging task.
Designing and building the physical components of CyPhA system involves knowledge of both electrical and civil engineering. 

In building the controller, e.g., in Stage-2, a major challenge is to construct a right set of rules for the controlling the water and the air pumps.  
Based on our background knowledge and experience in operating CyPhA system, we have kept [6.5, 8.5] and [3.5, 5.0]mg/L as the permissible ranges of pH and DO respectively. 
\begin{figure}
\begin{subfigure}{.32\linewidth}
\centerline{\includegraphics[width=\linewidth]{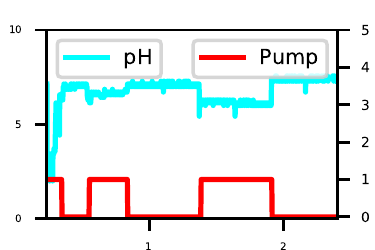}}
\label{fig}
\end{subfigure}
~
\begin{subfigure}{.32\linewidth}
\centerline{\includegraphics[width=\linewidth]{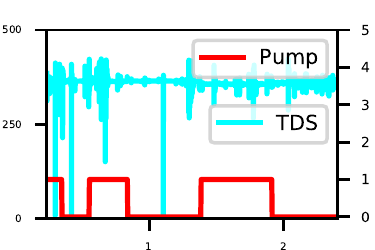}}
\label{fig}
\end{subfigure}
~
\begin{subfigure}{.32\linewidth}
\centerline{\includegraphics[width=\linewidth]{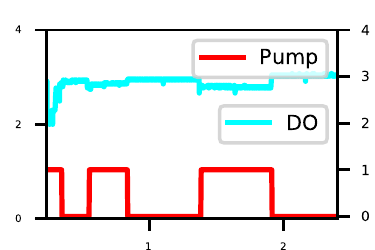}}
\label{fig}
\end{subfigure}
\caption{Regulate water pump (right-hand side y-axis) w.r.t. pH, TDS, DO (left-hand side y-axis) in Stage-2 on $30^{th}$ day.} 
\label{fig:water_param}
\end{figure}

Figure~\ref{fig:water_param} shows the duration for which the water pump in Stage-1 is switched on. 
For example, if pH goes below 6.5, the water pump in Stage-1 is switched on for a certain duration for transferring fresh water into the fish tank in Stage-2. 
Note that we pull out water from the fish tank using another pump before transferring water from Stage-1. 

\subsection{Future Work}
We aim to extend the work in CyPhA system by allowing remote users to connect over internet to access the system by developing appropriate security mechanisms. 
Also, we plan to extend the connectivity between CyPhA Gateway and the CyPhA controller from Wi-Fi to LoRa or similar technology to allow a larger geographic area for commercial application of the system. 

Another important extension of the work is to include analytics engine for real time data analysis for certain prediction and prescription in the aquaponics system, for example, we shall investigate how energy consumption can be reduced. 
Essentially, we shall focus on reducing operational cost through various ML techniques on the data collected out of CyPhA system. 

\section{Conclusion}
\label{sec:conclusion}
In this paper, we have designed and deployed a real testbed, called CyPhA system, for an aquaponics system that combines both fish farming and vegetable farming. 
The developed system is sustainable, remotely accessible and mimics a natural ecosystem where the fish eats the food and excrete waste, which is converted by beneficial bacteria to nutrients that are useful for vegetable farming. 
Open sourced tools, like ThingsBoard, are integrated with CyPhA system for data visualisation and providing user control commands. Data and command transfers use lightweight MQTT protocol across the modules. 
CyPhA system can be used for teaching, research and commercial applications.

\section*{Acknowledgment}

This work is partially supported by SERB grant (CRG/2020/005855).

%
%
%
\bibliographystyle{splncs04}
\bibliography{mybib}
\end{document}